# Emergence of New Materials for Exploiting Highly Efficient Carrier Multiplication in Photovoltaics

Sourav Maiti[§], Marco van der Laan[¥], Deepika Poonia[§], Peter Schall[¥], Sachin Kinge[§,#] and Laurens D.A. Siebbeles[§*]

[§]*Optoelectronic Materials Section, Department of Chemical Engineering, Delft University of Technology, Van der Maaseweg 9, Delft, 2629 HZ, The Netherlands*

[¥] *Institute of Physics, University of Amsterdam, 1098 XH Amsterdam, The Netherlands*

[#]*Toyota Motor Europe, Materials Research & Development, Hoge Wei 33, B-1913, Zaventem, Belgium*

Corresponding Authors: Sourav Maiti and Laurens D.A. Siebbeles

emails: s.maiti@tudelft.nl, l.d.a.siebbeles@tudelft.nl

**Abstract**

In conventional solar cell semiconductor materials (predominantly Si) photons with energy higher than the band gap initially generate hot electrons and holes, which subsequently cool down to the band edge by phonon emission. Due to the latter process, the energy of the charge carriers in excess of the band gap is lost as heat and does not contribute to the conversion of solar to electrical power. If the excess energy is more than the band gap it can in principle be utilized through a process known as carrier multiplication (CM) in which a single absorbed photon generates two (or more) pairs of electrons and holes. Thus, through CM the photon energy above twice the band gap enhances the photocurrent of a solar cell. In this review, we discuss recent progress in CM research in terms of fundamental understanding, emergence of new materials for efficient CM, and CM based solar cell applications. Based on our current understanding, the CM threshold can get close to the minimal value of twice the band gap in materials where a photon induces an asymmetric electronic transition from a deeper valence band or to a higher conduction band. In addition, the material must have a low exciton binding energy and high charge carrier mobility, so that photoexcitation leads directly to the formation of free charges that can readily be extracted at external electrodes of a photovoltaic device. Percolative networks of coupled PbSe quantum dots, Sn/Pb based halide perovskites, and transition metal dichalcogenides such as MoTe$_2$ fulfill these requirements to a large extent. These findings point towards promising prospects for further development of new materials for highly efficient photovoltaics.

Table of Contents

# 1. Introduction

A photon with energy hν exceeding the band gap ($E_g$) of a semiconductor can excite an electron from a valence band to a conduction band and create an electron-hole pair. In this way, a hot electron and hole are produced that usually thermalize quickly to the band-edge with the excess energy (hν − $E_g$) being lost as heat (Figure 1(a)). This poses a fundamental limitation to the efficiency of solar cells and one of the predominant reasons for the Shockley-Queisser limit of ~33% for single-junction solar cells.[1] Given sufficient excess energy, it can in principle be utilized to generate additional charge carriers through carrier multiplication (CM), as shown in Figure 1(b).[2-7] In this way, CM can enhance the photocurrent of a solar cell and help to surpass the Shockley-Queisser limit.[2,3,8,9]

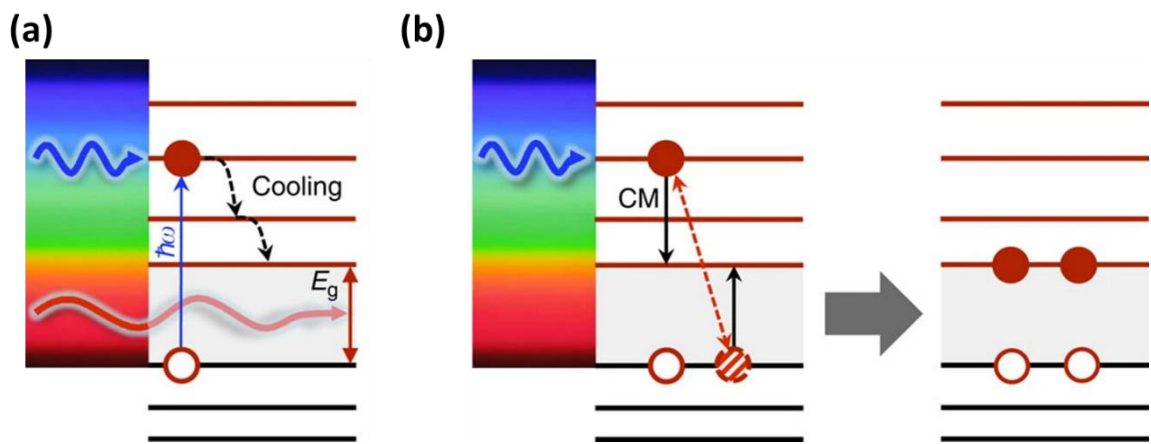

**Figure 1.** (a) Fast cooling of charge carriers leads to loss of excess energy for photoexcitations higher than the band gap; (b) In CM the excess energy of a carrier (electron here) is utilized for additional electron-hole pair generation. Reproduced with permission from ref. [10]. Copyright (2014) Macmillan Publishers Limited.

CM is also known as impact ionization (II) in bulk semiconductors and multi-exciton generation (MEG) in quantum confined nanomaterials when neutral excitons (Coulombically bound electron-hole pairs) are formed rather than free charge carriers. The key factors characterizing CM are the threshold photon energy from which CM starts and the quantum yield (QY), i.e. the number of electron-hole pairs produced per absorbed photon. The ideal CM scenario is a staircase dependence of the QY on the photon energy where the QY reaches 2 (n) at twice (n-times) the band gap multiple (Figure 2(a)). The band gap multiple is the photon energy normalized to the band gap of the material; i.e. $h\nu/E_g$.

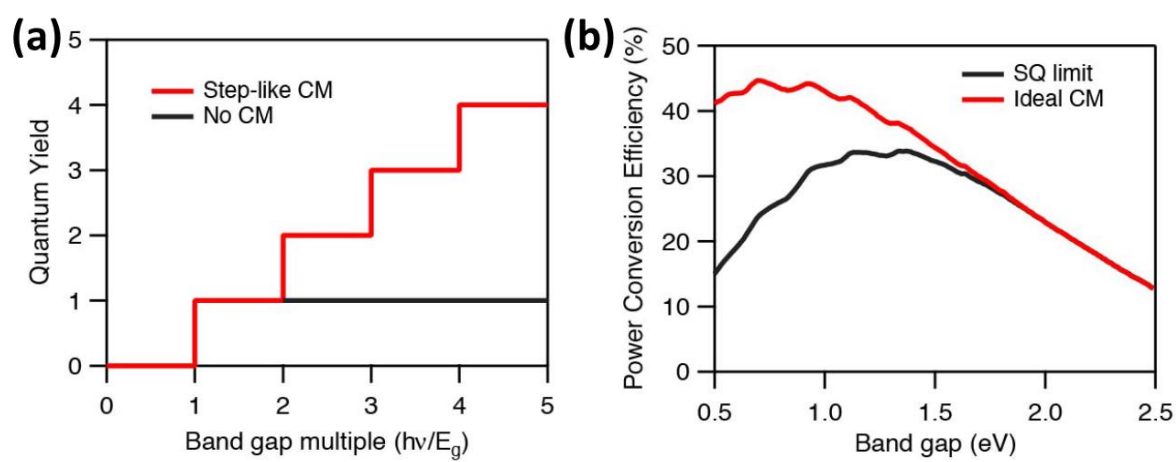

**Figure 2.** (a) The CM QY as a function of band gap multiple ($h\nu/E_g$) for ideal step-like CM; (b) the theoretical power conversion efficiency without (Shockley-Queisser limit) and with the ideal CM scenario. With kind permission from the ACS. The original article (ref. [11]) can be found at https://pubs.acs.org/doi/10.1021/acsnano.7b06511. Further permissions related to the material excerpted should be directed to the ACS.

To effectively exploit CM in solar cells the band gap of the semiconductor should be 0.6-1.0 eV resulting in a maximum efficiency of ~ 44% for an ideal staircase scenario, see Figure 2(b). Due to their suitable band gap (0.7-1.0 eV), Pb-chalcogenide based nanomaterials have been widely investigated for CM.[3,5,10-31] In addition, CM has also been studied in nanoparticles consisting of Cd-chalcogenides, Si, Ag$_2$S, CuInSe$_2$, as well as in 2D graphene and 1D carbon nanotubes.[12,32-44] Extensive reviews of advances in CM research have appeared, with recent ones by Pietryga et al.[45] in 2016, and by Kershaw et al.[46] in 2017. In this review, we describe the general understanding of CM and focus on recent research in the past three years. The latter includes studies of CM in Pb-chalcogenide heterostructures and networks, Si nanorods, perovskites, and transition metal dichalcogenides (TMDCs).[11,28,36,47-54] The current understanding of how the CM threshold is related to the band structure in terms of asymmetric optical excitations will be discussed in detail.[55] Recent results on the relatively high CM efficiencies found in weakly quantum confined and bulk perovskites and in TMDCs are of particular interest. Compared to quantum dots (QDs) the more facile charge transport in bulk perovskite and TMDCs are of particular interest for applications in photovoltaic devices. We include a brief discussion of CM-based solar cells and conclude with a future outlook.

## 2. Brief history of carrier multiplication

During the process of CM, a hot charge carrier with energy exceeding the band gap (either an electron in a conduction band or a hole in a valence band) relaxes by excitation of a valence band electron to the conduction band. CM occurs in competition with phonon emission (carrier cooling). In bulk materials the CM threshold is often as high as about 4 times the band gap.[40] In that case, CM is not useful for solar cell applications.

In 2002 Nozik theoretically proposed that CM in quantum confined nanomaterials can be more efficient than in bulk.[2] This sparked a lot of interest to study CM in semiconductor nanocrystals (NCs), especially in Pb-chalcogenide NCs because of their suitable band gap for solar cells. Soon after the work of Nozik, Schaller et al. reported experimental observation of efficient CM in PbSe QDs.[3] However, controversy arose with opposing results of the efficiency of CM in QDs from different laboratories.[32,56-60] Careful experimental procedures to avoid artifacts have shown the CM efficiency to be lower than the initial results in QDs, but still of promise for solar cell applications.[9,16,61] Later on the research of CM was extended to 1-D nanorods, 2-D nanosheets, complex heterostructures, and assemblies showing both a decrease of the CM threshold and an increase of the QY, see Section 6 of this paper. In the past few years efficient CM has been reported for (heterostructures of) Pb-chalcogenide based NCs of different shapes, with a CM threshold close to twice the band gap. More recently, efficient CM has also been observed in bulk perovskites and 2-D TMDCs. Interestingly, recent research suggests quantum confinement may not be a necessary requirement for efficient CM, as will be discussed in Section 6.

**3. Experimental techniques to investigate CM**

The experimental techniques mostly utilized to investigate CM involve time-resolved pump-probe laser spectroscopy with detection of transient optical absorption, photoluminescence, or microwave/terahertz conductivity.

**3. a. Transient optical absorption measurements**

Pump-probe transient optical absorption (TA) spectroscopy is the most widely used technique to characterize CM. In TA experiments, the sample is excited by a pump laser pulse creating electron-hole pairs (excitons or free charge carriers), which are probed by a time-delayed optical probe pulse to obtain the differential absorption ($\Delta A = A_{pump\ on} - A_{pump\ off} = \log(\frac{I_{off}^{probe}}{I_{on}^{probe}})$, where $A$ is the absorbance) as a function of time. A negative $\Delta A$ arises due to the depletion of the ground state population (ground state bleach, GSB) by the pump and/or stimulated emission from an exciton state. On the other hand, $\Delta A$ is positive if the photogenerated electron-hole pairs absorb the probe photons due to excitation to a higher state. The magnitude of $|\Delta A|$ normalized to the absorbed pump fluence ($I_0 F_A$, where $I_0$ is the incident number of photons per area and $F_A$ is the fraction absorbed) is directly proportional to the number of electron-hole pairs (with quantum yield $\varphi$):

$$\frac{|\Delta A|}{I_0 F_A} = \varphi \frac{\sigma_B}{\ln 10} \tag{1}$$

Here, $\sigma_B$ represents the cross-section of bleach, photoinduced absorption, and/or stimulated emission at the probe energy due to an electron-hole pair.

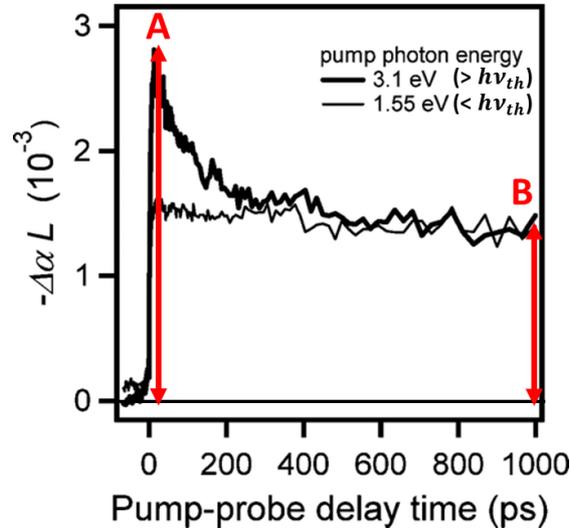

**Figure 3.** Absolute value of the ground state bleach ($-\Delta \alpha L = |\Delta A| \ln 10$) in PbSe QDs for below and above CM threshold photoexcitation with the same number of absorbed pump photons ($I_0 F_A$). The larger value of $|\Delta A|$ at early time (3 ps, red arrow at A) for the higher pump photon energy (3.1 eV) is indicative of CM. The subsequent fast decay component reflects Auger recombination of two or more CM generated excitons in the same QD. At longer times the magnitude of $|\Delta A|$ becomes the same as that for a pump photon energy of 1.55 eV

(red arrow at B), indicating that Auger recombination after 3.1 eV excitation is complete and $|\Delta A|$ is due to QDs containing a single exciton only. Adapted with permission from ref. [16]. Copyright (2008) American Chemical Society.

For sufficiently high pump photon energy the hot electrons and holes can undergo CM or cooling by phonon emission. Hot charge carriers can lead to another magnitude and shape of the TA spectrum than relaxed charges at the band gap.[62] To exclude such effects in the determination of the QY, the value of $|\Delta A|$ should be taken at a time when the hot carriers have relaxed and the spectral shape of the TA no longer varies with time. Then, for the same absorbed pump fluence, an increase of $|\Delta A|$ at higher pump photon energies indicates the occurrence of CM (Figure 3). After photogeneration of two or more excitons in a QD, the TA signal exhibits a rapid decay due to Auger recombination (Figure 3). Consequently, the TA signal on longer times is due to QDs containing one exciton only. In this case, the initial QY of excitons can be determined by taking the ratio of $|\Delta A|$ at an early time (A) when multi-excitons are still present and at a longer time (B) when the Auger process is complete leaving only one exciton in a QD (Figure 3).

## 3. b. Transient photoluminescence measurements

Transient photoluminescence (PL) measurements have also been utilized to determine the CM threshold and QY.[32,59,63] For pump photon energies below twice the band gap (and at sufficiently low fluence so that each QD absorbs at most one photon) the PL reflects the radiative decay of single excitons. A faster decay of the PL at higher pump photon energy (due to Auger recombination of multi-excitons in a QD) is indicative of CM.

## 3. c. Transient terahertz/microwave conductivity measurements

Free mobile charge carriers in assemblies of QDs, nanowires, 2-D, or bulk materials can be probed by time-resolved alternating current (AC) conductivity techniques at microwave or terahertz frequencies.[6,11,17,64,65] In the case of optical pump terahertz (THz) probe (OPTP) or microwave probe experiments the transient photoconductivity ($\Delta\sigma$) is obtained with picosecond and nanosecond time resolution, respectively. The magnitude of $\Delta\sigma$ is given by:

$$\Delta\sigma = eN_A\varphi(\mu_e + \mu_h) \quad . \tag{2}$$

Here $e$ is the elementary charge, $N_A$ is the number of absorbed pump photons per unit volume, $\varphi$ is the QY of charge carriers, and $\mu_e$ and $\mu_h$ are the electron and hole mobility, respectively. Therefore, the slope of the linear increase of $\Delta\sigma$ vs. $N_A$ gives the CM QY for the corresponding pump photon energy. If we excite below twice the band gap CM is impossible and the observed slope represents $\varphi = 1$. For excitation above twice the band gap an increase of the slope of a plot of $\Delta\sigma$ vs. $N_A$ gives the CM QY similar to the TA measurements discussed above.

With THz measurements, the magnitude of $\Delta\sigma$ can be obtained on a picosecond timescale, which in most cases is sufficiently short to ensure recombination or trapping of charges has not yet occurred. Microwave conductivity measurements have a time resolution of nanoseconds and recombination/trapping of electrons and holes may already have taken place. The latter may be slower at higher pump photon energies and therefore care must be taken that a higher photoconductivity on a nanosecond timescale does indeed reflect CM.

## 4. Factors affecting the CM threshold and efficiency

For an ideal case scenario, the CM threshold appears at twice the band gap with QY of 2. However, due to restrictions imposed by energy and momentum conservations the CM threshold is often far off from the ideal scenario. For parabolic bands with equal electron and hole effective masses, the threshold becomes 4 times the band gap as shown in Figure 4(a).[40] As momentum conservation rules are relaxed in QDs the CM threshold can be lower than for bulk material. In QDs with equal effective masses of electrons and holes, the CM threshold theoretically becomes 3 times the band gap Figure 4.(b)). Indeed, it has been shown experimentally that in Pb-chalcogenide QDs with almost equal effective masses of electrons and holes the CM threshold is close to thrice the band gap (Figure 5).[8,9,55] The QY increases almost linearly above the threshold and the steeper the slope the higher is the CM efficiency.

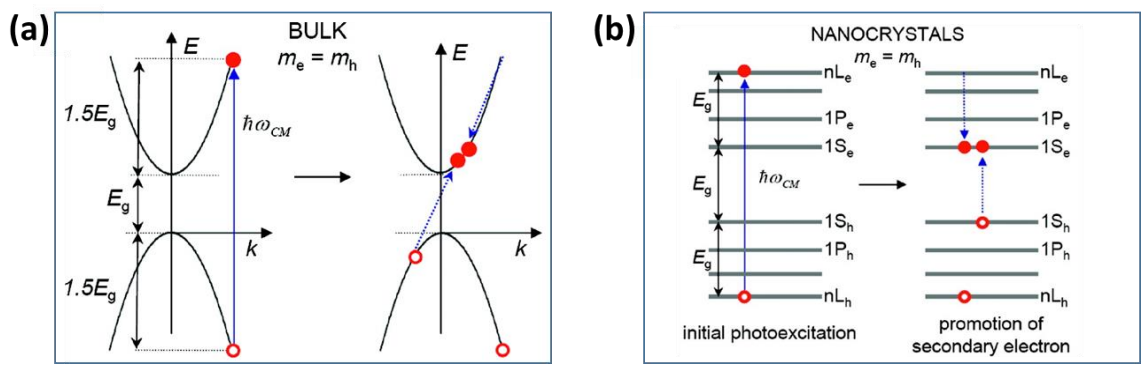

**Figure 4.** (a) Due to energy and momentum conservation rules the CM threshold in bulk can be as high as 4 times the band gap for equal electron and hole effective masses. (b) In the case of QDs, the relaxation of momentum conservation rules can result in a much lower CM threshold. Reproduced with permission from ref. [40]. Copyright (2007) American Chemical Society.

In the context of solar cell applications, the CM QY is usually plotted *vs.* the photon energy normalized to the band gap, which is denoted as the band gap multiple, defined by $h\nu/E_g$. The CM efficiency ($\eta_{CM}$) is defined as the change of the QY with the change of the band gap multiple $h\nu/E_g$ according to[9]:

$$\eta_{CM} = \frac{\Delta(QY)}{\Delta(\frac{h\nu}{E_g})} \quad (3)$$

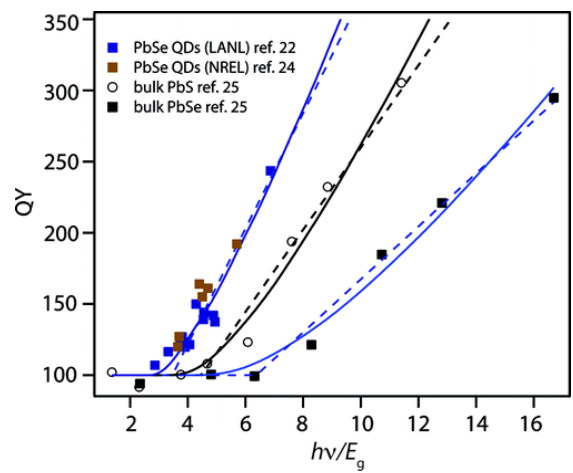

**Figure 5.** CM QY *vs.* band gap multiple ($h\nu/E_g$) for PbSe and PbS QDs in comparison with the bulk. The dashed lines have been obtained by fits of Equation (3) to the experimental data. Reproduced with permission from ref. [9]. Copyright (2010) American Chemical Society.

## 4. a. The CM threshold is related to asymmetric optical excitations

If the excess photon energy above the band gap is almost entirely transferred to either the electron or the hole the CM threshold can be near twice the band gap. Such asymmetric photoexcitation is possible if the effective mass of the electron and hole are largely different, which is the case for InAs QDs ($m_e/m_h \sim 0.05$).[40] In this case the excess photon energy is almost completely transferred to the electron and the CM threshold is close to twice the band gap.[40]

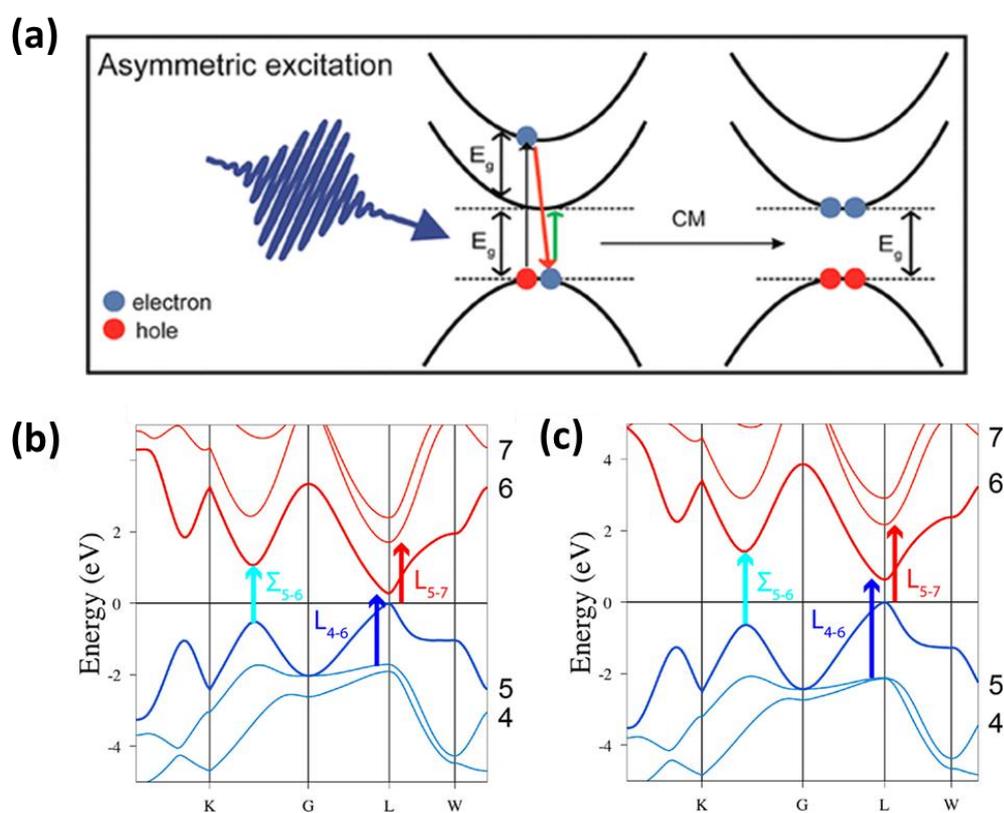

**Figure 6.** (a) The concept of asymmetric photoexcitation involves an unequal distribution of the excess photon energy an electron and a hole. Photogeneration of the carrier with most of the excess energy (here electron) determines the CM threshold. With kind permission from the ACS. The original article (ref. [11]) can be found at https://pubs.acs.org/doi/10.1021/acsnano.7b06511. Further permissions related to the material excerpted should be directed to the ACS. Bulk band structure of (b) PbSe and (c) PbS with the asymmetric $L_{4-6}$ and $L_{5-7}$ excitations indicated. With kind permission from the ACS. The original article (ref. [55]) can be found at https://pubs.acs.org/doi/10.1021/acsnano.8b01530. Further permissions related to the material excerpted should be directed to the ACS.

Asymmetric photoexcitation as mentioned above is also possible if there is a second conduction (or valence) band with extremum at twice the band gap, as shown in Figure 6(a). In that case, the excess photon energy is fully transferred to the electron (or hole), which can subsequently relax by CM. Recently, Spoor *et al.* have shown that for Pb-chalcogenide bulk and QDs the CM threshold can be correlated to the onset of asymmetric optical excitations.[55] These asymmetric excitations involve higher valence and conduction bands, as shown in Figure 6(b) and 6(c). Here, the $L_{5-6}$ excitation is across the band gap, whereas the $L_{5-7}$ and $L_{4-6}$ excitations refer to transitions from the highest valence band to the 2$^{nd}$ conduction band and the 2$^{nd}$ valence band to the lowest conduction band, respectively. Figure 7 shows the CM QY and the threshold as a function of the band gap energy together with $L_{5-7}$ and $L_{4-6}$ excitation energies (normalized to the band gap energy) for different sizes of PbSe and PbS QDs and their bulk counterparts. The graphs show an excellent correlation between the CM threshold and the $L_{5-7}$ and $L_{4-6}$ excitation energies. For photon energies at which the $L_{5-7}$ and $L_{4-6}$ are possible an asymmetric excitation can cause either the electron ($L_{5-7}$) or the hole ($L_{4-6}$) to carry most of the excess energy. The finding that the CM threshold is close to the onset of asymmetric excitations implies that at this energy CM outcompetes carrier cooling.[55] Hence, quantum confinement may not be a strict requirement for CM as long as a second conduction or valence band exists and CM outcompetes carrier cooling.

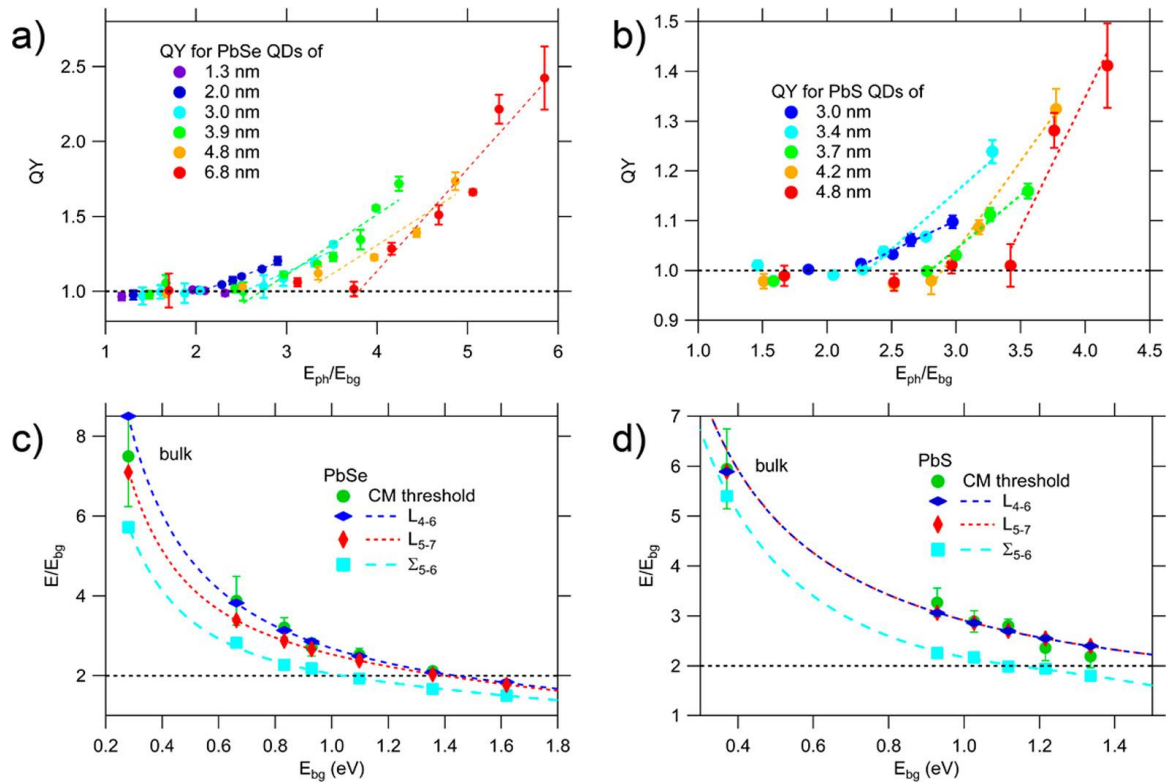

**Figure 7.** QY *vs.* band gap multiple in case of (a) PbSe and (b) PbS QDs of different sizes. (c,d) Band gap multiple $E/E_{bg}=h\nu/E_g$ as a function of the band gap ($E_{bg}$) showing the relation between the calculated asymmetric L$_{4-6}$ and L$_{5-7}$ excitations and the CM threshold for different sizes of PbSe (c) and PbS QDs (d). With kind permission from the ACS. The original article (ref. [55]) can be found at https://pubs.acs.org/doi/10.1021/acsnano.8b01530. Further permissions related to the material excerpted should be directed to the ACS.

## 5. Theory of carrier multiplication

The CM QY is the net result of the decay of a hot charge carrier *via* consecutive steps of CM and phonon emission. CM results from the coupling of single and multi-exciton states by Coulomb interactions. In the context of quantum chemistry, this is known as configuration interaction (CI) of excited Slater determinants within the Born-Oppenheimer approximation for fixed nuclei.[66,67] A coherent superposition of single and multi-exciton states for fixed nuclei is hypothetical, due to the coupling of electrons with nuclear lattice vibrations (phonons), which makes the Born-Oppenheimer approximation invalid. Electron-phonon coupling results in electron cooling to lower states by phonon emission, as well as decoherence. Numerical calculations of incoherent decay of initially energetic charge carriers, with the rates of the competitive CM and phonon emission processes as parameters, have been successfully used to reproduce CM QYs in PbSe and PbS QDs.[9,55]

To date, the theoretical description of the rate of CM on the basis of Coulomb coupling between single and multi-exciton sates in NCs has predominantly focused on the formation of biexcitons and at most triexcitons.[68] Higher order CM processes have not been treated yet due to the large computational effort needed. The theoretical work started about 15 years ago with the introduction of three classes of CM pathways, reviewed in refs. [45,46] and briefly described below.

Firstly, the pathway in the model of Shabaev *et al.*[30,69] considers photoexcitation from the electronic ground state to an excited state that is a coherent superposition of a single exciton state and a biexciton state. The coupling between the single and biexciton states is assumed to be coherent due to strong Coulomb interaction. The coherent excited state can relax to uncoupled lower lying single exciton and biexciton states with a phenomenological rate that describes phonon emission. This gives rise to depopulation and dephasing of the single and biexciton states in the initially photogenerated coherent state. Pure dephasing is not taken into account. In a later work the model was extended to a full quantum-state evolution approach to describe CM in PbSe QDs, including the formation of triexcitons.[68]

However, direct comparison with experimental results was hindered by the fact that taking into account all exciton states was computationally too demanding.

A second approach known as the direct photogeneration model assumes a weak Coulomb coupling between single and biexciton states and was introduced by Klimov *et al.*[70,71] Photoexcitation is assumed to occur at an energy that is resonant with a biexciton state, but off-resonant with single exciton states. One mechanism involves an off-resonant 'virtual' single exciton state making photoexcitation from the ground state to the biexciton possible *via* an optical dipole transition.[71] This mechanism was used to explain the ultrashort timescale of CM in CdSe and PbSe QDs. Another mechanism corresponds to a ground state that is a mixture of a state with no excitons (vacuum state) coupled to a biexciton state by Coulomb interaction (this is analogous to CI in quantum chemistry).[70] The admixture of the biexciton in the ground state allows direct photoexcitation to a higher biexciton state that is resonant with the photon energy.

A third class of CM pathways has been introduced by Zunger *et al.*,[72] Delerue *et al.*,[73], and Rabani and Baer.[74] Their models are based on Fermi's golden rule to calculate the rate of CM due to the Coulomb coupling between an initially photogenerated single exciton and final biexciton states. The CM QY can be obtained by including the optical oscillator strength for photoexcitation to single exciton states and their subsequent competitive decay *via* CM and phonon emission. Phonon emission has been treated as a single step process with the rate being an adjustable parameter. It has been found that the final density of states (DOS) of biexcitons decreases as NCs become smaller and this effect by itself reduces the CM rate. This does not imply that the CM rate decreases for smaller NCs, since the enhanced Coulomb interaction in smaller NCs can compensate the effect of a reduced DOS. The impact ionization model has been applied to describe CM in a variety of NCs with electronic states obtained from *e.g.* atomistic tight-binding, pseudopotential, density functional theory calculations, or **k·p** theory.[19,55,68,75-77]

A very general theoretical treatment of CM has been provided by Piryatinski and Velizhanin and is known as the exciton scattering model.[78] This model is applicable to cases ranging from weak to strong Coulomb coupling and includes the above described CM pathways as limiting cases. The exciton scattering model is applicable to NCs and bulk and takes into account the photoexcitation dynamics of an electron by an optical pulse, and its subsequent relaxation *via* CM in competition with phonon emission. The general approach is realized by integrating scattering theory in the time propagation of the system, which is described by the density matrix formalism. The only restriction of the exciton scattering model is that it does not include states with multiplicity higher than biexcitons. The model has been used to numerically analyze experimental CM QYs in PbSe and PbS QDs and bulk and it was found that the impact ionization is the predominant pathway involved in CM.[79,80]

*Ab initio* time-domain studies of the evolution of a photoexcited electronic state have been carried out by Prezhdo and coworkers.[66] These are first-principle calculations without invoking phonon relaxation rates as parameters. The methodology involves propagation of the time-dependent electronic wavefunction consisting of the ground state and single and biexciton states.[81] Non-adiabatic effects of the motion of the nuclei on the evolution of the electronic wave function are explicitly taken into account. The nuclear motion in the potential of the electrons is described classically. These numerical calculations are computationally very demanding and could thus far be applied to clusters consisting of at most tens of atoms. However, the results give qualitative insights that are useful to analyze the effects of material composition, phonons, and structural defects on CM and Auger recombination of electrons and holes.

## 6. Emerging materials for efficient CM

### 6. a. Pb-chalcogenide 1-D nanorods and 2-D nanosheets

The characteristics of CM in 1-D Pb-chalcogenide nanorods (NRs) and 2-D nanosheets (NSs) differ from that in their 0-D QD counterparts. The CM QY in Pb-chalcogenide NRs with aspect ratio near 6 is about two times higher than for PbSe QDs with a similar band gap (Figure 8.(a)).[23] The better performance of NRs can be due to enhanced Coulomb interaction between charge carriers resulting from electric field lines penetrating through the low dielectric medium surrounding the NRs. Interestingly, the Auger decay lifetime was longer in PbSe NRs than in QDs, which is beneficial for charge extraction in a solar cell.

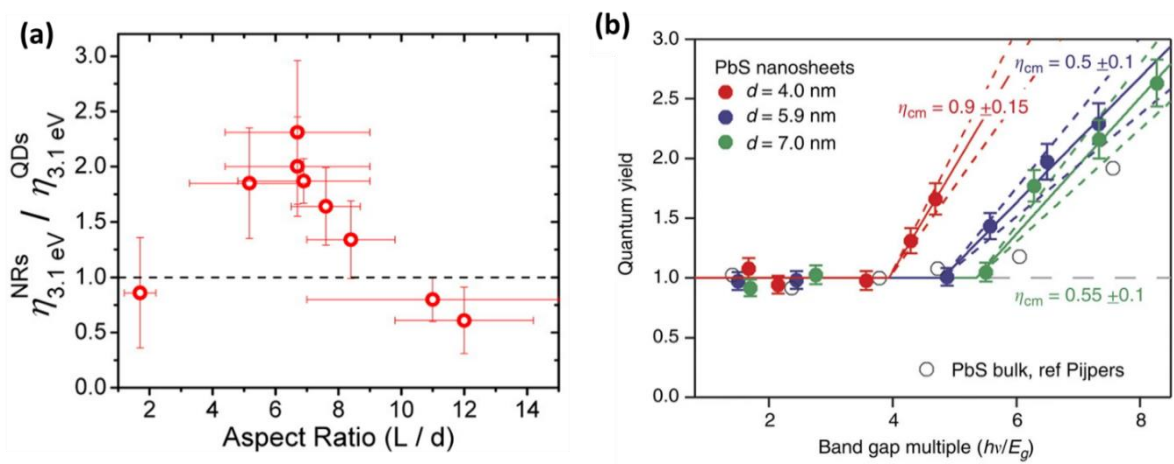

**Figure 8.** (a) The ratio of the CM QY for PbSe QDs and 1-D NRs of similar band gap. The plot shows that NRs with aspect ratio 6-7 have a higher CM QY than QDs. Reprinted with permission from ref. [23]. Copyright (2013) American Chemical Society. (b) CM QY *vs.* band gap

multiple for PbS nanosheets with thicknesses as indicated. The slope of the linear fits is CM efficiency ($\eta_{CM}$). Reprinted with permission from ref. [27]. Copyright (2014) Macmillan Publishers Limited.

For NSs of PbS, it has been found that their thickness drastically influences the CM threshold and QY, see Figure 8.(b). For 4 nm thick NSs the CM threshold is near 4 times the band gap, while the CM efficiency ($\eta_{CM}$) is close to 1. Hence, above the CM threshold the excess photon energy is almost fully utilized to generate additional carriers by CM. The CM efficiency in NSs is higher than for PbS QDs and bulk. As the thickness increases the CM threshold becomes higher and the CM efficiency decreases, see Figure 8.(b). Due to the high CM threshold, the 4 nm thick NSs are not of interest for solar cell applications. However, the observed reduction of the CM threshold as the thickness decreases makes it of interest to study if this trend continues for thinner PbS NSs, while maintaining a CM efficiency close to unity.

**6. b. Nanocrystal heterostructures**

The usual symmetric optical excitations in Pb-chalcogenides can be made asymmetric in a heterostructure with a Cd-chalcogenide. This is possible due to the almost equal energy of the conduction band of these two materials, while the valence band of Cd-chalcogenides is lower in energy than for Pb-chalcogenides. Asymmetric excitations have been realized in core/shell QDs and Janus-like NCs, as discussed below.

**i. Core/shell quantum dots**

Asymmetric optical excitation has been demonstrated for core/shell PbSe/CdSe QDs and a CM threshold close to twice the band gap (~2.2 $E_g$) has been realized.[10] The CM QY was found to be higher than for PbSe NRs of a similar band gap (Figure 9). Core/shell QDs have several properties that are beneficial to CM: (i) the PbSe core and the CdSe shell share a common conduction band, but the valence band offset is 1.48 eV. This causes the hole to be strongly confined in the core, increasing the hole energy level spacing which can slow down the cooling rate. (ii) For photon energies more than twice the band gap the optical excitations mainly involve electrons from the CdSe-shell, which is due to the higher absorption cross-section of CdSe. Hence, above twice the band gap the hole is created in the CdSe-shell dominated state, which makes the optical excitation asymmetric with the hole having most of the excess energy. This leads to a CM threshold just above twice the band gap. These QDs also exhibit a higher CM QY due to the slow rate of hole cooling, resulting from the low density of hole states in the PbSe core. Indeed, the hot hole emission lifetime is as long as 6-10 picoseconds, which corroborates that cooling is much slower than CM.

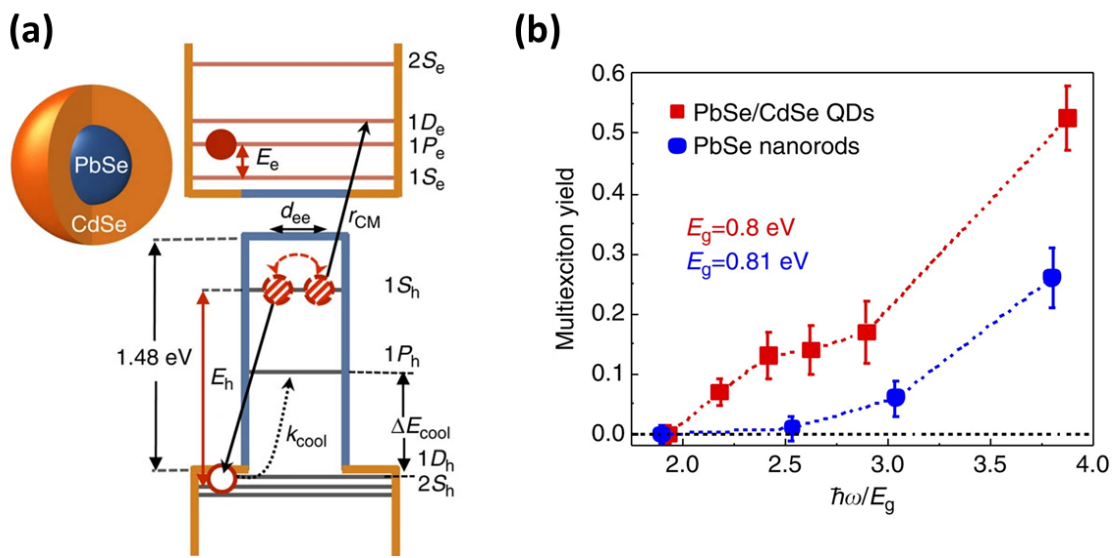

**Figure 9.** (a) Electronic energy levels in a PbSe/CdSe core/shell QD. The PbSe and CdSe share the same conduction band energies, which are distributed throughout the QD. In the valence band, the energy levels are sparsely distributed in the PbSe core with separation of $\Delta E_{cool}$. After photoexcitation the excess energy is distributed asymmetrically with the electron excess energy ($E_e$) being much smaller than that of the hole ($E_h$). The hole cooling rate ($k_{cool}$, dotted

black arrow) is much slower than CM (solid black arrow). (b) Multiexciton yield (number of excitons above one; *i.e.* QY-1) *vs.* band gap multiple for PbSe/CdSe QDs and PbSe NRs of similar band gap (0.8 eV) showing more efficient CM in PbSe/CdSe QDs. The threshold in the QDs is close to twice the band gap, due to the asymmetric distribution of excess photon energy. Reprinted with permission from ref. [10]. Copyright (2014) Macmillan Publishers Limited.

**ii. Janus heterostructures**

Kroupa *et al.* have shown that CdS/PbS Janus hetero-structures have a CM threshold close to twice the band gap (which is determined by the PbS component) and QY higher than core/shell QDs.[28] The Janus structure allows asymmetric optical excitations, see Figure 10. It was theoretically estimated that ~25% of the optical excitations above the CM threshold create hot holes with more excess energy than the electron. The holes get trapped at interfacial states at the CdS/PbS heterojunction within 1 picosecond and undergo CM rather than cooling by phonon emission (Figure 10). Note, that in the PbSe/CdSe core/shell QDs discussed above the hole is confined in the PbSe core and is difficult to extract. Reverse core/shell CdSe/PbSe would be ideal for charge extraction but are difficult to synthesize. In this regard, Janus structures where both charge carriers are accessible from the NC surface are promising candidates for photovoltaics. However, the difficulty is to deposit the Janus NCs so that all the CdS (and PbS) are selectively connected together so that the electron (hole) can move from one particle to another with ease and finally gets extracted at the electron (hole) contacts.

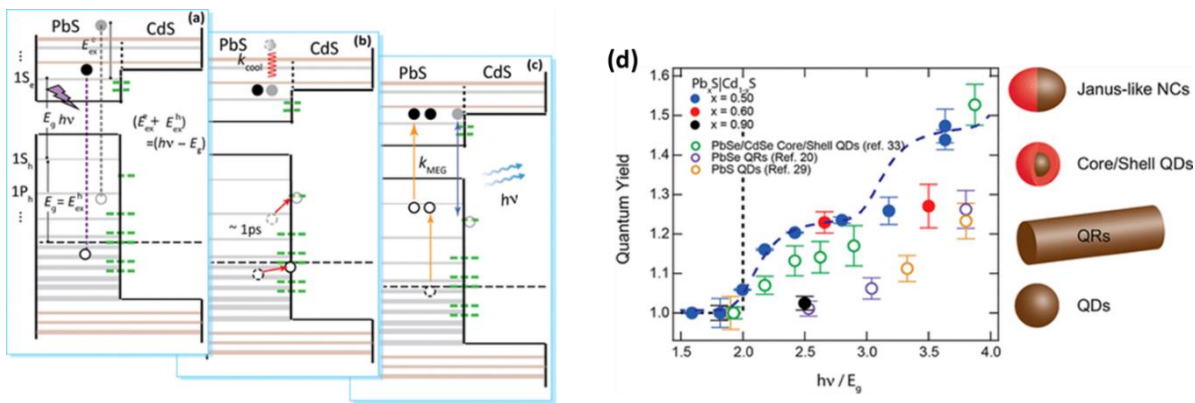

**Figure 10.** (a) Absorption of a photon with energy just above twice the band gap can result in an asymmetric excitation with the valence band hole having most of the excess energy (black circles). More symmetric excitations are also possible as shown by the grey circles. (b) A hot hole can get trapped at interfacial states (dashed green lines) within 1 ps. (c) The trapped holes can undergo CM to generate additional charge carriers. (d) CM QY *vs.* band gap multiple for Janus CdS/PbS NCs, PbSe/CdSe core/shell QDs, PbSe NRs, and PbS QDs. It is observed that the CM threshold is close to twice the band gap for both PbSe/CdSe core/shell QDs and Janus CdS/PbS NCs, due to asymmetric optical excitations. Reprinted with permission from ref. [28]. Copyright (2018) American Chemical Society.

**6. c. Pb-chalcogenide networks**

We have discussed the individual Pb-chalcogenide NCs (QDs, NRs, NSs) and heterostructures (core/shell, Janus) in terms of CM threshold and QY. In heterostructures, the CM threshold is reduced to just above twice the band gap and the QY is higher than in NCs consisting of a Pb-chalcogenide only. However, for photovoltaic device applications, the NCs must be coupled to allow charge carrier transport and extraction at external electrodes. This can be realized by mutually connecting NCs to form an assembly in which charges can move from one NC to another. Therefore, from a practical point of view characterization of CM in solid films of coupled NCs is essential.

In the first instance, Pb-chalcogenide QDs were coupled by introducing short organic ligands on their surface or infilling the space between QDs with metaloxides.[13,17,25,26,82] While this yielded encouraging results, a breakthrough in terms of a low CM threshold and relatively high QY was realized by Kulkarni *et al.* in a percolative PbSe NC network with a band gap of 0.7 eV, which is suitable to exploit CM in a solar cell.[11] In this network, the original QDs are directly connected *via* strong crystalline PbSe bridges.[83,84] The efficiency of CM was studied using OPTP spectroscopy, see Section 3.c. Figure 11(a) shows that the THz conductivity increases with photoexcitation energy at twice the band gap. Interestingly, a stepwise behavior was found for the QY *vs.* the band gap multiple (Figure 11(b)), which has never been observed for uncoupled QDs in dispersion. The low CM threshold must be due to an asymmetric excitation where the excess energy ends up solely either in the electron or the hole. If a 2$^{nd}$ VB or CB exists close to twice the band gap then a CM threshold at this energy is possible, as discussed in Section 4. Electronic structure calculations on percolative networks are needed to corroborate the occurrence of such asymmetric transitions.

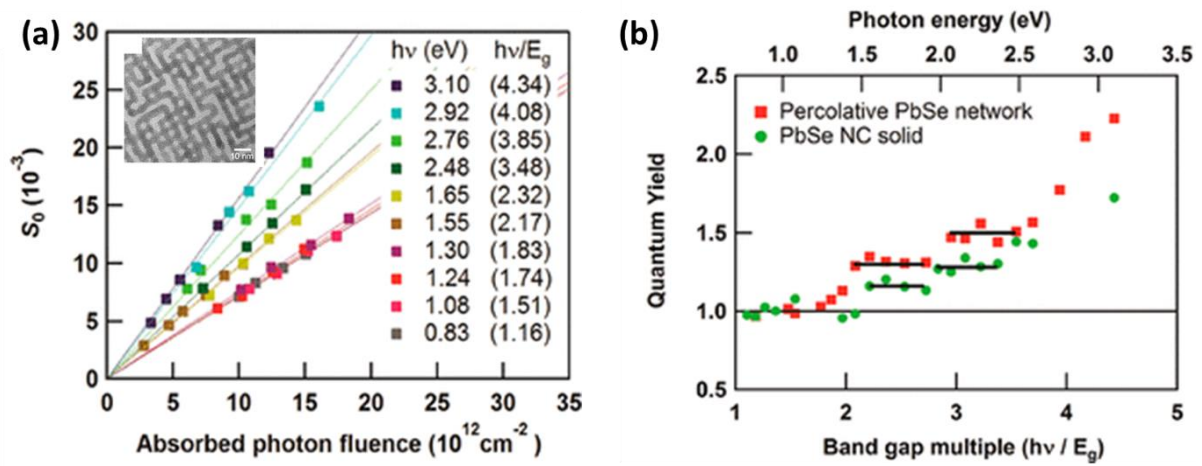

**Figure 11.** (a) The THz photoconductivity signal ($S_0$) of a percolative PbSe network (structure in inset) at short time (2 ps) after the pump laser pulse as a function absorbed photon fluence for below and above twice the band gap (0.7 eV). The solid lines represent linear fits and the increase of the slope above twice the band gap indicates CM. (b) CM QY *vs.* the band gap multiple for the percolative PbSe network and a PbSe NC solid coupled by organic ligands (1,2 ethanediamine). The QY for the percolative PbSe network is significantly higher than for the NC solid with organic ligands. With kind permission from ACS. The original article (ref. [11]) can be found at https://pubs.acs.org/doi/10.1021/acsnano.7b06511. Further permissions related to the material excerpted should be directed to the ACS.

**6. d. Si nanorods**

In bulk Si, the predominant semiconductor utilized in solar cells, the CM threshold is about 3.5 times the band gap and the QY becomes 140% at 4.5 times the band gap.[85] Therefore, CM in bulk Si is not useful for practical applications. However, Si NCs can be attractive candidates for multi-excitonic solar cells due to quantum confinement effects. CM in Si QDs (9.5 nm diameter and band gap of 1.2 eV) was shown to have a threshold of 2.4 times the band gap and the CM QY is 2.6 at 3.4 times the band gap.[34] Recently, CM in Si NRs with three different aspect ratios (diameter 7.5 nm, aspect ratio around 6, 20, and 33) and band gap ~ 1.3 eV has been reported by Stolle *et al.* through TA measurements, see Figure 12.[36] The photoinduced absorption (PIA) of Si in the NIR region (~1200 nm) was monitored at different photoexcitation energies. For excitation energies more than twice the band gap the increase in $|\Delta A|$ with a biexcitonic Auger decay confirmed the CM process. The CM threshold is lower for Si NRs of aspect ratio 20 (2.2 times the band gap) than it is for Si QDs (2.6 times the band gap) with similar band gap.

Importantly, the CM QY was found to be 1.6 at 2.9 times the band gap which is twice that of Si QDs (Figure 12).

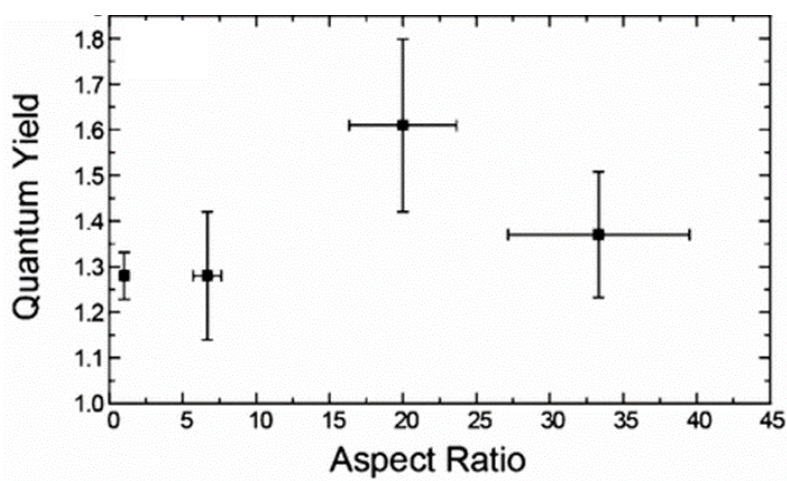

**Figure 12.** CM QY at 3.86 eV photoexcitation for Si QDs (aspect ratio 1, the first data point) and Si NRs with different aspect ratios. Reprinted with permission from ref. [36] Copyright (2017) American Chemical Society.

**6. e. Perovskite materials**

CM in perovskite materials has been investigated for several compositions such as in organic-inorganic halide perovskite formamidinium lead iodide (FAPbI$_3$), all-inorganic cesium lead iodide (CsPbI$_3$) and bulk Sn/Pb halide perovskites.[47-50]

**i. FAPbI$_3$ NCs**

Li *et al.* studied the CM efficiency in cubic FAPbI$_3$ perovskite NCs of different sizes and in a bulk sample, which have a varying degree of quantum confinement.[47] CM was characterized through a fast decay of the TA signal due to biexciton Auger recombination at pump photon energies higher than twice the band gap, see Figure 13. CM was not observed in the bulk film and was negligible (1.07 ± 0.05 at 2.94Eg) for the weakly confined size of 12.9 m. For the intermediate confined size (7.5 nm) the CM QY was found to be 1.32 ± 0.06 at hv = 2.7Eg with the CM threshold at 2.25 times the band gap. The CM QY was found to be increasing linearly with the photon energy above the threshold with a slope of 0.75 for the 7.5 nm NCs (Figure 13). The CM performance of these perovskite NCs is better than for Pb-chalcogenide QDs in terms of a lower CM threshold and a higher efficiency. The superior performance was explained in terms of slower charge carrier cooling and strong Coulomb interactions.

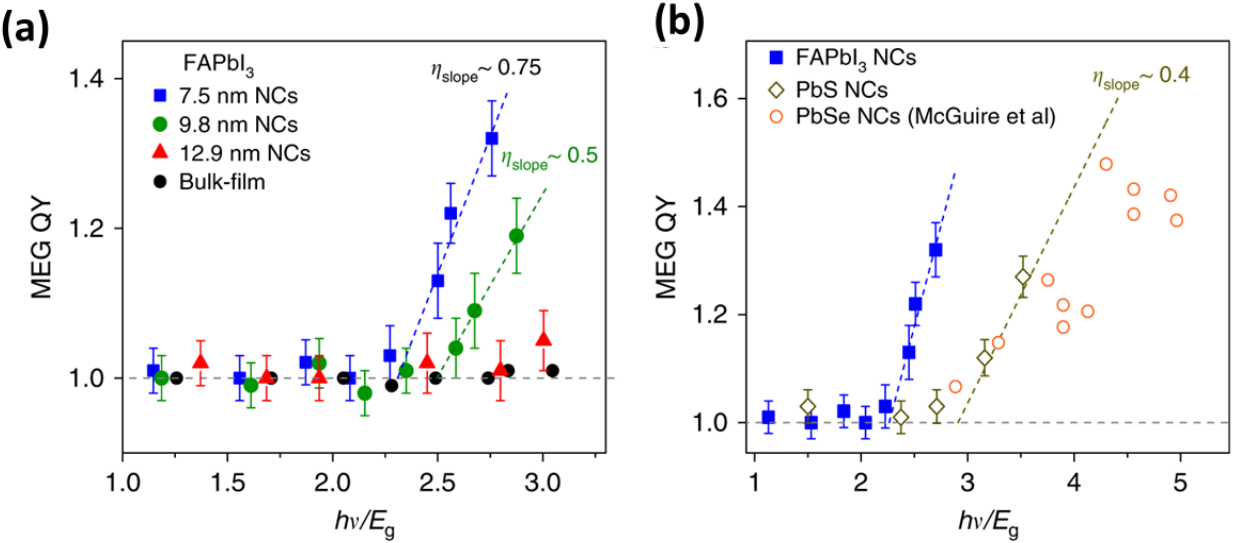

**Figure 13.** (a) CM QY (denoted as MEG QY in the figure) vs. band gap multiple for cubic FAPbI$_3$ NCs of 7.5 nm (intermediate confinement), 9.8 nm, and 12.9 nm (weak confinement) together with data for bulk. The dashed lines are linear fits to the data with the slope representing the CM efficiency ($\eta_{CM}$). (b) Comparison of the CM QY of the intermediate confined FAPbI$_3$ NCs with PbS and PbSe QDs, showing a lower CM threshold and higher CM efficiency in the perovskite NCs. Adapted with permission from ref. [47].

**ii. CsPbI$_3$ NCs**

CM in cubic CsPbI$_3$ NCs has been investigated by several groups. de Weerd *et al.* reported efficient CM in very weakly confined CsPbI$_3$ NCs of 11.5 nm size through TA measurements.[48] The CM QY is shown as a function of band gap multiple in Figure 14. The CM threshold is close to twice the band gap with CM efficiency near unity. Interestingly, Makarov *et al*. did not observe any CM (measured through ultrafast PL decay) in these cubic CsPbI$_3$ perovskite NCs, which were synthesized according to the same protocol as used by the Weerd *et al*.[86] The difference was attributed to subtle differences in the surface structure and local stoichiometry. Recently, Cong *et al*. reported CM in CsPbI$_3$ NCs in the strong confinement region, but found CM to be insignificant in the weak confinement region.[49] They attributed the more significant CM in smaller NCs to stronger Coulomb interactions. The varying results from different groups require additional studies to understand the factors that govern CM in CsPbI$_3$ NCs. It should be noted that these perovskites are not suitable for solar cell applications due to their high band gap.

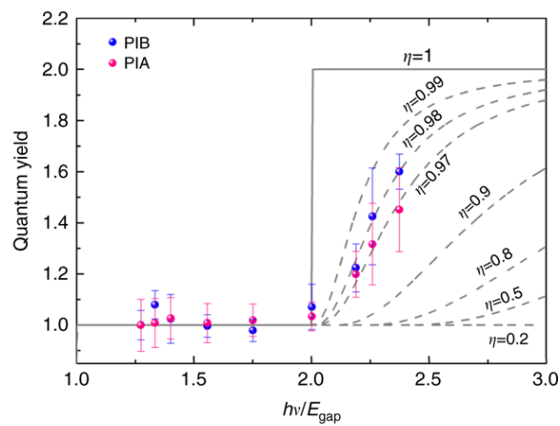

**Figure 14.** CM QY *vs.* the band gap multiple for cubic CsPbI$_3$ NCs showing a CM threshold close to twice the band gap. Reproduced with permission from ref. [48].

### iii. Sn/Pb halide perovskites

Mixed Sn/Pb halide perovskites have a band gap as low as 1.28 eV, which is much more suitable for solar cell applications than the band gap of the perovskites discussed above.[50] Recently, Maiti *et al.* have shown efficient CM in a bulk Sn/Pb halide perovskite of the composition $(FASnI_3)_{0.6}(MAPbI_3)_{0.4}$. The CM threshold was found to be just above twice the band gap and the QY reaches 2 at 2.8 times the band gap (Figure 15).[50] Asymmetric excitation in which the excess photon energy is transferred to the electron, is a plausible explanation for the low CM threshold and high QY, as a recent theoretical study showed the presence of a second conduction band close to 2.2 times the band gap.[87] The mixed Sn/Pb halide perovskite has a low exciton binding energy (~16 meV), so that photoexcitation will predominantly lead to the generation of free charges at room temperature, which is important for photovoltaic applications. Also, the bulk structure is better for charge transport than assemblies of NCs.

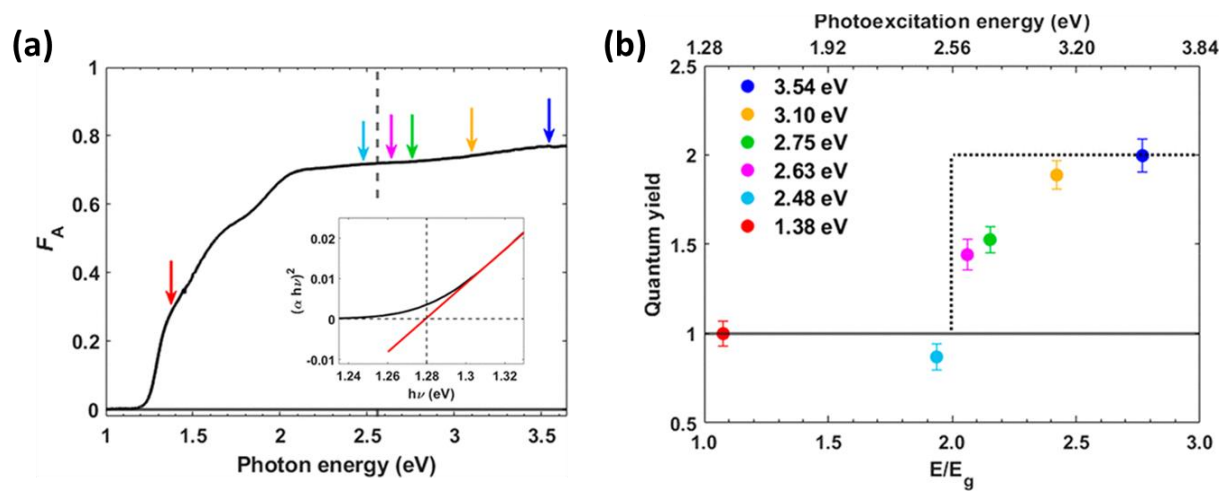

**Figure 15.** (a) Absorption spectrum of the mixed Sn/Pb halide perovskite with a low band gap of 1.28 eV. (Inset) Tauc plot to determine the band gap. (b) CM QY *vs.* band gap multiple showing a CM threshold close to twice the band gap and a QY reaching 2 at 2.8 times the band gap. With kind permission from the ACS. The original article (ref. [50]) can be found at

### 6. f. Transition metal dichalcogenides

Semiconducting 2-D transition metal dichalcogenides, $MX_2$ (M = transition metal, X = chalcogen), are currently receiving attention for the exploitation of CM in photovoltaics. Recently, two studies have been published on CM in $MoTe_2$ investigated by pump-probe experiments.[51,52] The latest work by Zheng *et al.* involves CM in $MoTe_2$ (5 nm thick, ~7 layers, indirect band gap 0.90 eV), studied using OPTP spectroscopy (Section 3.c).[52] The CM threshold was 2.8 times the band gap with an ideal CM QY of 2 and showing a staircase like behavior with the QY reaching 3 at 4.2 times the band gap (Figure 16(a)). To explain the CM characteristics asymmetric optical excitations were invoked (Figure 16(b)). Photoexcitation across the indirect band gap of $MoTe_2$ involves a K-Λ transition, whereas near the CM threshold photoexcitation can occur *via* a direct transition at the Γ point. As the K and Γ points have similar valence band energy, the excess photon energy is mostly transferred to the electron. The hot electron can relax in the conduction band *via* the Γ-Λ transition and produce another electron-hole pair through K-Λ excitation. Moreover, it was argued that weak electron-phonon coupling in $MoTe_2$ reduces the loss of excess photon energy by charge carrier cooling.

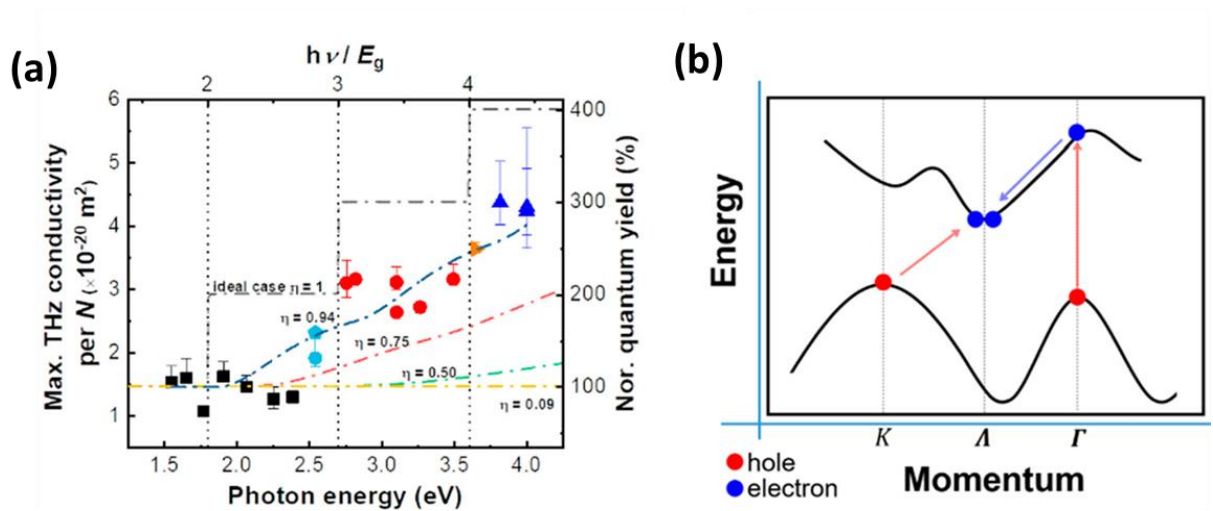

**Figure 16.** (a) Maximum THz conductivity *vs.* band gap multiple in MoTe$_2$ showing CM threshold around 3 times the band gap. (b) Schematic of asymmetric excitation in MoTe$_2$. With kind permission from the ACS. The original article (ref. [52]) can be found at https://pubs.acs.org/doi/10.1021/acs.nanolett.0c01693. Further permissions related to the material excerpted should be directed to the ACS.

These results are slightly different from the earlier ones reported by Kim *et al.* who employed TA to investigate CM in MoTe$_2$ (16.5 nm thick film) with an indirect bandgap of 0.85 eV.[51] CM was characterized by TA measurements, as outlined in Section 3.a. They reported a CM threshold close to twice the band gap and QY reaching 2 at 2.7 times the band gap (Figure 17).

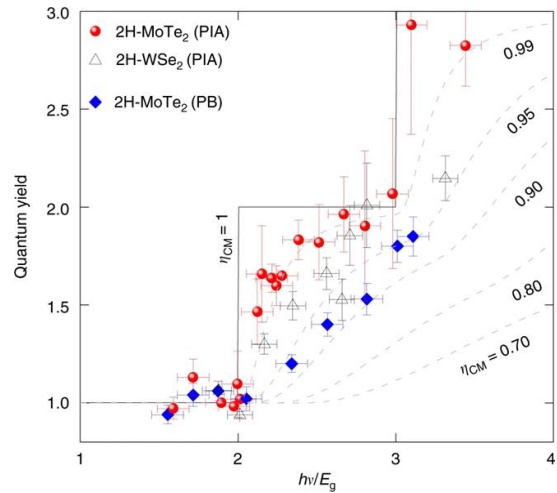

**Figure 17.** CM QY vs. band gap multiple in MoTe$_2$ and WSe$_2$. Reprinted with permission from ref. [51].

The reason for the difference in the results by Zheng *et al*.[52] and Kim *et al*.[51] is not clear at the moment. Some degree of variation in the number of defects and the doping level could be due to different sample preparation procedures. Future studies are needed to shed light on the role of sample morphology and to eliminate defects to enhance the charge carrier lifetime.[52] However, the current results make MoTe$_2$ of interest for further studies directed towards photovoltaic applications. In addition, for all molybdenum and tungsten based TMDCs, material thickness seems to mostly affect the indirect transition,[88] leaving the direct transition at the K-point relatively stable, which potentially allows for new ways of tuning the asymmetric excitations necessary for efficient CM.

## 7. Carrier multiplication in photovoltaic device applications

It can be concluded from pump-probe spectroscopy, that in several materials CM occurs efficiently with a threshold close to twice the band gap. However, studies showing the enhancement of photocurrent due to CM in photovoltaic devices are limited due to the difficulty of realizing efficient transport of charge carriers and their extraction at the electrodes. Proof-of-concept solar cells with internal quantum efficiency exceeding 100% have been reported for PbS QDs attached to $TiO_2$ by a mercaptopropionic acid linker.[89] The fast extraction of the electron (~ 50 fs) to $TiO_2$ and hole (~4 ps) by a polysulphide electrolyte ensures efficient charge extraction before Auger recombination or trapping. An external quantum efficiency (EQE) exceeding 100% has been reported for solar cells based on PbSe QDs[90], $CuInS_2$ QDs[39], PbSe NRs[91], or PbTe QDs[92]. However, for Janus PbS/CdS heterostructure NCs (Section 6.b) solar cells have not shown an external quantum efficiency higher than 100%, despite the fact that CM was observed by TA measurements.[28] Therefore, research is required to improve the device architecture for fast charge carrier transport and efficient extraction at the electrodes in a solar cell.[93]

Recently, Kim *et al*. have developed a conductive atomic force microscope (CAFM) system to measure the local photocurrent in PbS QDs (5.4 nm diameter) for different photon energies.[94] The photocurrent was measured between an Au tip decorated with PbS QDs and a graphene layer on a $SiO_2$/Si substrate. Interestingly, a step-like CM behavior was found with a threshold close to twice the band gap and near-ideal CM efficiency (Figure 18). The advantage of this method is that it probes the local current between the QD and Au tip so that charge transport between QDs does not play a role.

Barati *et al.* investigated CM in a TMDC heterostructure consisting of MoSe$_2$ and WSe$_2$.[54] Both in photocurrent and $I_{SD}$ - $V_G$ measurements CM was found to occur with QY up to 3.5. In this case, CM is due to an impact ionization-like process induced by the applied source-drain voltage.

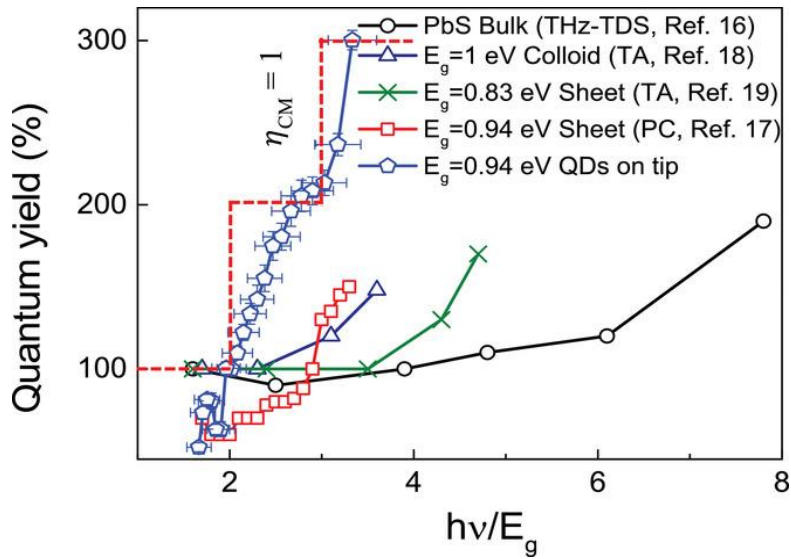

**Figure 18.** The CM QY as a function of band gap multiple for PbS QD coated Au tips in comparison with PbS bulk, QDs, and nanosheets. Reproduced with permission from ref. [94]. Copyright (2019) Macmillan Publishers Limited. Copyright (2020) WILEY-VCH Verlag GmbH & Co.

## 8. Conclusion and future outlook

We have discussed recent advances of research on CM and findings of new materials exhibiting near-ideal CM. It is of interest that CM with low threshold and appreciable QY has been found to occur in two-dimensional and bulk materials. Conditions to be met for the significant impact of CM in photovoltaics are: (i) asymmetric photoexcitation in which the excess photon energy is transferred predominantly to the electron or the hole so that the CM threshold can be close to twice the band gap, (ii) the exciton binding energy must be sufficiently small to generate free charge carriers, and (iii) charge carrier mobilities need to be high enough for efficient charge carrier transport and collection at electrodes in a device. These conditions have been realized to a large extent in percolative PbSe networks, a bulk Sn/Pb halide perovskite, and MoTe$_2$. It appears that quantum confinement is not strictly required for efficient CM. Therefore, future research should also focus on two-dimensional and bulk-like materials.

**Acknowledgments**

This research received funding from the Netherlands Organisation for Scientific Research (NWO) in the framework of the Materials for Sustainability Programme (MAT4SUS project numbers 739.017.001 and 739.017.011) and from the Ministry of Economic Affairs in the framework of the PPP allowance.

**Data availability statement**

Data sharing is not applicable to this review article as no new data were created or analyzed.

**References**

1    W. Shockley and H. J. Queisser, J. Appl. Phys. 32, 510 (1961).

2    A. J. Nozik, Physica E: Low-dimensional Systems and Nanostructures 14, 115 (2002).

3    R. D. Schaller and V. I. Klimov, Phys. Rev. Lett. 92, 186601 (2004).

4    J. A. McGuire, J. Joo, J. M. Pietryga, R. D. Schaller, and V. I. Klimov, Acc. Chem. Res. 41, 1810 (2008).

5   L. A. Padilha, J. T. Stewart, R. L. Sandberg, W. K. Bae, W.-K. Koh, J. M. Pietryga, and V. I. Klimov, Acc. Chem. Res. 46, 1261 (2013).

6   S. ten Cate, C. S. S. Sandeep, Y. Liu, M. Law, S. Kinge, A. J. Houtepen, J. M. Schins, and L. D. A. Siebbeles, Acc. Chem. Res. 48, 174 (2015).

7   M. S. Martinez, A. J. Nozik, and M. C. Beard, J. Chem. Phys. 151, 114111 (2019).

8   M. C. Beard, J. M. Luther, O. E. Semonin, and A. J. Nozik, Acc. Chem. Res. 46, 1252 (2013).

9   M. C. Beard, A. G. Midgett, M. C. Hanna, J. M. Luther, B. K. Hughes, and A. J. Nozik, Nano Lett. 10, 3019 (2010).

10  C. M. Cirloganu, L. A. Padilha, Q. Lin, N. S. Makarov, K. A. Velizhanin, H. Luo, I. Robel, J. M. Pietryga, and V. I. Klimov, Nat. Commun. 5, 4148 (2014).

11  A. Kulkarni, W. H. Evers, S. Tomić, M. C. Beard, D. Vanmaekelbergh, and L. D. A. Siebbeles, ACS Nano 12, 378 (2018).

12  R. D. Schaller, M. A. Petruska, and V. I. Klimov, Appl. Phys. Lett. 87, 253102 (2005).

13  R. J. Ellingson, M. C. Beard, J. C. Johnson, P. Yu, O. I. Micic, A. J. Nozik, A. Shabaev, and A. L. Efros, Nano Lett. 5, 865 (2005).

14  J. E. Murphy, M. C. Beard, A. G. Norman, S. P. Ahrenkiel, J. C. Johnson, P. Yu, O. I. Mićić, R. J. Ellingson, and A. J. Nozik, J. Am. Chem. Soc. 128, 3241 (2006).

15  J. M. Luther, M. C. Beard, Q. Song, M. Law, R. J. Ellingson, and A. J. Nozik, Nano Lett. 7, 1779 (2007).

16  M. T. Trinh, A. J. Houtepen, J. M. Schins, T. Hanrath, J. Piris, W. Knulst, A. P. Goossens, and L. D. Siebbeles, Nano Lett. 8, 1713 (2008).

17  M. Aerts, C. S. Suchand Sandeep, Y. Gao, T. J. Savenije, J. M. Schins, A. J. Houtepen, S. Kinge, and L. D. A. Siebbeles, Nano Lett. 11, 4485 (2011).

[18]   G. Nootz, L. A. Padilha, L. Levina, V. Sukhovatkin, S. Webster, L. Brzozowski, E. H. Sargent, D. J. Hagan, and E. W. Van Stryland,  Phys. Rev. B 83, 155302 (2011).

[19]   M. T. Trinh, L. Polak, J. M. Schins, A. J. Houtepen, R. Vaxenburg, G. I. Maikov, G. Grinbom, A. G. Midgett, J. M. Luther, M. C. Beard, A. J. Nozik, M. Bonn, E. Lifshitz, and L. D. A. Siebbeles,  Nano Lett. 11, 1623 (2011).

[20]   P. D. Cunningham, J. E. Boercker, E. E. Foos, M. P. Lumb, A. R. Smith, J. G. Tischler, and J. S. Melinger,  Nano Lett. 11, 3476 (2011).

[21]   P. D. Cunningham, J. E. Boercker, E. E. Foos, M. P. Lumb, A. R. Smith, J. G. Tischler, and J. S. Melinger,  Nano Lett. 13, 3003 (2013).

[22]   A. G. Midgett, J. M. Luther, J. T. Stewart, D. K. Smith, L. A. Padilha, V. I. Klimov, A. J. Nozik, and M. C. Beard,  Nano Lett. 13, 3078 (2013).

[23]   L. A. Padilha, J. T. Stewart, R. L. Sandberg, W. K. Bae, W.-K. Koh, J. M. Pietryga, and V. I. Klimov,  Nano Lett. 13, 1092 (2013).

[24]   S. ten Cate, Y. Liu, J. M. Schins, M. Law, and L. D. A. Siebbeles,  J. Phys. Chem. Lett. 4, 3257 (2013).

[25]   S. ten Cate, Y. Liu, C. S. Suchand Sandeep, S. Kinge, A. J. Houtepen, T. J. Savenije, J. M. Schins, M. Law, and L. D. A. Siebbeles,  J. Phys. Chem. Lett. 4, 1766 (2013).

[26]   C. S. S. Sandeep, S. ten Cate, J. M. Schins, T. J. Savenije, Y. Liu, M. Law, S. Kinge, A. J. Houtepen, and L. D. A. Siebbeles,  Nat. Commun. 4, 2360 (2013).

[27]   M. Aerts, T. Bielewicz, C. Klinke, F. C. Grozema, A. J. Houtepen, J. M. Schins, and L. D. A. Siebbeles,  Nat. Commun. 5, 3789 (2014).

[28]   D. M. Kroupa, G. F. Pach, M. Vörös, F. Giberti, B. D. Chernomordik, R. W. Crisp, A. J. Nozik, J. C. Johnson, R. Singh, V. I. Klimov, G. Galli, and M. C. Beard,  ACS Nano 12, 10084 (2018).

29   F. C. M. Spoor, G. Grimaldi, S. Kinge, A. J. Houtepen, and L. D. A. Siebbeles, ACS Appl. Energy Mater. 2, 721 (2019).

30   A. Shabaev, C. S. Hellberg, and A. L. Efros, Acc. Chem. Res. 46, 1242 (2013).

31   C. Smith and D. Binks, Nanomaterials 4, 19 (2014).

32   G. Nair and M. G. Bawendi, Phys. Rev. B 76, 081304 (2007).

33   D. Gachet, A. Avidan, I. Pinkas, and D. Oron, Nano Lett. 10, 164 (2010).

34   M. C. Beard, K. P. Knutsen, P. Yu, J. M. Luther, Q. Song, W. K. Metzger, R. J. Ellingson, and A. J. Nozik, Nano Lett. 7, 2506 (2007).

35   D. Timmerman, J. Valenta, K. Dohnalová, W. D. A. M. de Boer, and T. Gregorkiewicz, Nat. Nanotechnol. 6, 710 (2011).

36   C. J. Stolle, X. Lu, Y. Yu, R. D. Schaller, and B. A. Korgel, Nano Lett. 17, 5580 (2017).

37   J. Sun, W. Yu, A. Usman, T. T. Isimjan, S. Dgobbo, E. Alarousu, K. Takanabe, and O. F. Mohammed, J. Phys. Chem. Lett. 5, 659 (2014).

38   C. J. Stolle, R. D. Schaller, and B. A. Korgel, J. Phys. Chem. Lett. 5, 3169 (2014).

39   C. J. Stolle, T. B. Harvey, D. R. Pernik, J. I. Hibbert, J. Du, D. J. Rhee, V. A. Akhavan, R. D. Schaller, and B. A. Korgel, J. Phys. Chem. Lett. 5, 304 (2014).

40   R. D. Schaller, J. M. Pietryga, and V. I. Klimov, Nano Lett. 7, 3469 (2007).

41   K. J. Tielrooij, J. C. W. Song, S. A. Jensen, A. Centeno, A. Pesquera, A. Zurutuza Elorza, M. Bonn, L. S. Levitov, and F. H. L. Koppens, Nat. Phys. 9, 248 (2013).

42   N. M. Gabor, Z. Zhong, K. Bosnick, J. Park, and P. L. McEuen, Science 325, 1367 (2009).

43   S. Wang, M. Khafizov, X. Tu, M. Zheng, and T. D. Krauss, Nano Lett. 10, 2381 (2010).

44   N. Siemons and A. Serafini, J. Nanotechnol. 2018, 7285483 (2018).

45   J. M. Pietryga, Y. S. Park, J. H. Lim, A. F. Fidler, W. K. Bae, S. Brovelli, and V. I. Klimov, Chem. Rev. 116, 10513 (2016).

46    S. V. Kershaw and A. L. Rogach, Materials (Basel) 10, 1095 (2017).

47    M. Li, R. Begum, J. Fu, Q. Xu, T. M. Koh, S. A. Veldhuis, M. Grätzel, N. Mathews, S. Mhaisalkar, and T. C. Sum, Nat. Commun. 9, 4197 (2018).

48    C. de Weerd, L. Gomez, A. Capretti, D. M. Lebrun, E. Matsubara, J. Lin, M. Ashida, F. C. M. Spoor, L. D. A. Siebbeles, A. J. Houtepen, K. Suenaga, Y. Fujiwara, and T. Gregorkiewicz, Nat. Commun. 9, 4199 (2018).

49    M. Cong, B. Yang, J. Chen, F. Hong, S. Yang, W. Deng, and K. Han, J. Phys. Chem. Lett. 11, 1921 (2020).

50    S. Maiti, S. Ferro, D. Poonia, B. Ehrler, S. Kinge, and L. D. A. Siebbeles, J. Phys. Chem. Lett., 6146 (2020).

51    J.-H. Kim, M. R. Bergren, J. C. Park, S. Adhikari, M. Lorke, T. Frauenheim, D.-H. Choe, B. Kim, H. Choi, T. Gregorkiewicz, and Y. H. Lee, Nat. Commun. 10, 5488 (2019).

52    W. Zheng, M. Bonn, and H. I. Wang, Nano Lett. (2020).

53    A. Manzi, Y. Tong, J. Feucht, E.-P. Yao, L. Polavarapu, A. S. Urban, and J. Feldmann, Nat. Commun. 9, 1518 (2018).

54    F. Barati, M. Grossnickle, S. Su, R. K. Lake, V. Aji, and N. M. Gabor, Nat. Nanotechnol. 12, 1134 (2017).

55    F. C. M. Spoor, G. Grimaldi, C. Delerue, W. H. Evers, R. W. Crisp, P. Geiregat, Z. Hens, A. J. Houtepen, and L. D. A. Siebbeles, ACS Nano 12, 4796 (2018).

56    M. Ben-Lulu, D. Mocatta, M. Bonn, U. Banin, and S. Ruhman, Nano Lett. 8, 1207 (2008).

57    J. J. H. Pijpers, E. Hendry, M. T. W. Milder, R. Fanciulli, J. Savolainen, J. L. Herek, D. Vanmaekelbergh, S. Ruhman, D. Mocatta, D. Oron, A. Aharoni, U. Banin, and M. Bonn, J. Phys. Chem. C 112, 4783 (2008).

58    G. Nair, L.-Y. Chang, S. M. Geyer, and M. G. Bawendi, Nano Lett. 11, 2145 (2011).

59      G. Nair, S. M. Geyer, L.-Y. Chang, and M. G. Bawendi,  Phys. Rev. B 78, 125325 (2008).

60      I. Gdor, H. Sachs, A. Roitblat, D. B. Strasfeld, M. G. Bawendi, and S. Ruhman,  ACS Nano 6, 3269 (2012).

61      J. A. McGuire, M. Sykora, J. Joo, J. M. Pietryga, and V. I. Klimov,  Nano Lett. 10, 2049 (2010).

62      S. Schmitt-Rink, D. S. Chemla, and D. A. B. Miller,  Phys. Rev. B 32, 6601 (1985).

63      R. D. Schaller, M. Sykora, S. Jeong, and V. I. Klimov,  The Journal of Physical Chemistry B 110, 25332 (2006).

64      J. J. H. Pijpers, R. Ulbricht, K. J. Tielrooij, A. Osherov, Y. Golan, C. Delerue, G. Allan, and M. Bonn,  Nat. Phys. 5, 811 (2009).

65      J. Lauth, S. Kinge, and L. D.A. Siebbeles,  Z. Phys. Chem. 231, 107 (2017).

66      H. M. Jaeger, K. Hyeon-Deuk, and O. V. Prezhdo,  Acc. Chem. Res. 46, 1280 (2013).

67      F. Jensen, *Introduction to Computational Chemistry*. (John Wiley & Sons, New York, 2017).

68      W. M. Witzel, A. Shabaev, C. S. Hellberg, V. L. Jacobs, and A. L. Efros,  Phys. Rev. Lett. 105, 137401 (2010).

69      A. Shabaev, A. L. Efros, and A. J. Nozik,  Nano Lett. 6, 2856 (2006).

70      V. I. Rupasov and V. I. Klimov,  Phys. Rev. B 76, 125321 (2007).

71      R. D. Schaller, V. M. Agranovich, and V. I. Klimov,  Nat. Phys. 1, 189 (2005).

72      A. Franceschetti, J. M. An, and A. Zunger,  Nano Lett. 6, 2191 (2006).

73      G. Allan and C. Delerue,  Phys. Rev. B 73, 205423 (2006).

74      E. Rabani and R. Baer,  Nano Lett. 8, 4488 (2008).

75      J. W. Luo, A. Franceschetti, and A. Zunger,  Nano Lett. 8, 3174 (2008).

76      G. Zohar, R. Baer, and E. Rabani,  J Phys Chem Lett 4, 317 (2013).

77   H. Eshet, R. Baer, D. Neuhauser, and E. Rabani,  Nat. Commun. 7, 13178 (2016).

78   A. Piryatinski and K. A. Velizhanin,  J. Chem. Phys. 133, 084508 (2010).

79   K. A. Velizhanin and A. Piryatinski,  Phys. Rev. Lett. 106, 207401 (2011).

80   K. A. Velizhanin and A. Piryatinski,  Phys. Rev. B 86, 165319 (2012).

81   K. Hyeon-Deuk and O. V. Prezhdo,  ACS Nano 6, 1239 (2012).

82   G. W. Guglietta, B. T. Diroll, E. A. Gaulding, J. L. Fordham, S. M. Li, C. B. Murray, and J. B. Baxter,  ACS Nano 9, 1820 (2015).

83   M. P. Boneschanscher, W. H. Evers, J. J. Geuchies, T. Altantzis, B. Goris, F. T. Rabouw, S. A. P. van Rossum, H. S. J. van der Zant, L. D. A. Siebbeles, G. Van Tendeloo, I. Swart, J. Hilhorst, A. V. Petukhov, S. Bals, and D. Vanmaekelbergh,  Science 344, 1377 (2014).

84   W. H. Evers, J. M. Schins, M. Aerts, A. Kulkarni, P. Capiod, M. Berthe, B. Grandidier, C. Delerue, H. S. J. van der Zant, C. van Overbeek, J. L. Peters, D. Vanmaekelbergh, and L. D. A. Siebbeles,  Nat. Commun. 6, 8195 (2015).

85   M. Wolf, R. Brendel, J. H. Werner, and H. J. Queisser,  J. Appl. Phys. 83, 4213 (1998).

86   N. S. Makarov, S. Guo, O. Isaienko, W. Liu, I. Robel, and V. I. Klimov,  Nano Lett. 16, 2349 (2016).

87   L. Guan, X. Xu, Y. Liang, S. Han, J. Guo, J. Wang, and X. Li,  Phys. Lett. A 384, 126173 (2020).

88   W. S. Yun, S. W. Han, S. C. Hong, I. G. Kim, and J. D. Lee,  Phys. Rev. B 85, 033305 (2012).

89   J. B. Sambur, T. Novet, and B. A. Parkinson,  Science 330, 63 (2010).

90   O. E. Semonin, J. M. Luther, S. Choi, H.-Y. Chen, J. Gao, A. J. Nozik, and M. C. Beard,  Science 334, 1530 (2011).

91   N. J. L. K. Davis, M. L. Böhm, M. Tabachnyk, F. Wisnivesky-Rocca-Rivarola, T. C. Jellicoe, C. Ducati, B. Ehrler, and N. C. Greenham,  Nat. Commun. 6, 8259 (2015).

[92] M. L. Böhm, T. C. Jellicoe, M. Tabachnyk, N. J. L. K. Davis, F. Wisnivesky-Rocca-Rivarola, C. Ducati, B. Ehrler, A. A. Bakulin, and N. C. Greenham, Nano Lett. 15, 7987 (2015).

[93] H. Goodwin, T. C. Jellicoe, N. J. L. K. Davis, and M. L. Böhm, Nanophotonics 7, 111 (2018).

[94] S.-T. Kim, J.-H. Kim, and Y. H. Lee, Adv. Mater. 32, 1908461 (2020).